\documentclass[preprint2]{aastex}

\slugcomment{Submitted to Astrophysical Journal Letters}
\shortauthors{H.\ Krawczynski}
\shorttitle{Spectroscopic Deprojection-Analysis of 3C~129}
\begin{document}
\title{
Spectroscopic Deprojection-Analysis of Chandra Data of the 
Galaxy~Cluster~3C~129}
\author{H. Krawczynski\altaffilmark{1}}
\altaffiltext{1}{Yale University, P.O. Box 208101, New Haven, CT 06520-8101, USA, email: krawcz\@astro.yale.edu}
\begin{abstract}
We report on spectroscopic imaging observations of the nearby ($z$=0.022)  
galaxy cluster 3C~129 performed with the ACIS detector on board of the Chandra
X-ray observatory. Applying a deprojection analysis which fully takes into
account the spatially resolved X-ray energy spectra we investigate the 
radial variation of temperature and particle density of the Intracluster 
Medium (ICM).
While earlier data indicated a weak cooling flow and a cool cluster core, 
the Chandra data suggests that the core is hotter than the rest of the 
cluster: at the center we infer 7.9$^{+4.4}_{-2.4}$~keV while the mean 
temperature of all the other ICM shells is 5.1$\pm$0.3~keV 
(all errors on 90\% confidence level).
Based on the radial entropy profile we discuss the evidence for shock 
heating of the ICM at the cluster center. 
We discuss the possibility to use imaging spectroscopy data of more regular 
clusters to extract maps of the gravitational dark matter potential.
\end{abstract}
\section{Introduction}
\label{intro}
While observations with the ROSAT and ASCA X-ray telescopes revealed
non-homogeneous temperature of the ICM for a number of clusters
(see e.g.\ Honda et al.\ 1996), the large scale radial temperature 
structure remained uncertain due to ROSAT's limited spectral resolution 
and coverage, and ASCA's non-negligible energy dependent point spread function.
Based on ASCA observations of 30 galaxy clusters, Markevitch et al.\ (1998)
found evidence for a general temperature decline with distance from 
cluster center. However, independent ASCA studies of the same and other
galaxy clusters did not confirm this result (e.g., White 2000, 
Kikuchi et al. 1999).
The new X-ray observatories Chandra and XMM-Newton with broad energy coverage
of 0.3-10~keV (Chandra) and 0.1-15~keV (XMM-Newton) and angular
resolutions of 0.5~arcsec (Chandra) and 15~arcsec (XMM-Newton) make it now
possible to assess spatially resolved ICM X-ray energy spectra with
sufficiently broad energy coverage and excellent angular resolutions.
First observations of temperature or metallicity variations have already been
reported for a number of clusters, e.g.\ Coma (Arnaud et al.\ 2000), A~2142
(Markevitch et al.\ 2000), Hydra~A (McNamara et al.\ 2000), A~1795 
(Tamura et al.\ 2001), S\'{e}rsic 159-03 (Kaastra et al.\ 2001), and
Abell 1835 (Schmidt et al.\ 2001).
On theoretical grounds one expects that the formation of clusters 
gives rise to substantial temperature, entropy, and metallicity gradients.
The interplay of gravitational collapse and heating by accretion 
shocks produces characteristic radial temperature and entropy profiles
(e.g.\ Evrard 1990). Depending on when the ICM is enriched 
with metals, also radial metallicity gradients could be observed
\cite{Metz:94}. Furthermore, merger of major cluster components 
produce pronounced ICM non-uniformities, and hydro/N-body simulations predict
that the gradients could be observed several $10^9$ years after the onset
of the merger event \cite{Roet:96}.
The decay of these non-uniformities might be reduced by magnetic fields
which organize the ICM into a filamentary structure and thereby 
substantially reduce the ICM's heat conductivity and particle
diffusion (Eilek 1999, but see also Narayan \& Medvedev 2001).

In this letter we describe Chandra observations of the 
nearby rich cluster of galaxies 3C~129.
Due to its low galactic latitude ($l$=160$^\circ$.5, $b$=0.3$^\circ$)
the cluster has not been studied intensively at optical wavelengths.
Based on ROSAT, Einstein, and EXOSAT data, Leahy \& Yin (2000) 
estimated a total 0.2-10~keV luminosity of 
2.7$\times10^{44}$~ergs s$^{-1}$, a total ICM gas mass of 
3.6$\times10^{13}$~M$_{\odot}$, and a total cluster mass of
$\sim$5$\,\times\,10^{14}$~M$_{\odot}$. The cluster harbors two radio 
galaxies: the prototypical head-tail galaxy 3C~129, and the 
weaker FR~I source 3C~129.1.
The main emphasis of the Chandra observation had been a study of the
pressure balance between the ICM and the radio plasma and this study will
be presented elsewhere.

In this letter we present a ``spectroscopic deprojection-analysis'' 
of the Chandra data.
The deprojection-approach has been the standard-technique to convert 
observed X-ray surface brightness profiles into temperature 
and particle density profiles (see Fabian 1994, and references therein).
Lacking spatially resolved X-ray energy spectra, these studies
used several assumptions, namely that the ICM had a certain geometry,
that the ICM was in hydrostatic equilibrium, and that 
the dark matter gravitational potential could be described by
a certain shape and depth, together with an ICM temperature and 
metal abundance estimate derived from a non-imaging X-ray observation.
The spatially resolved energy spectra from the Chandra and XMM-Newton 
observatories now make it possible to directly derive the radial temperature, 
particle density and metallicity profiles based only on assumptions
about the ICM geometry. A similar analysis has independently been 
developed by Allen et al.\ (2001). 

The rest of this letter is organized as follows.
After describing the data set in Sect.~\ref{data} and the 
spectroscopic deprojection-method in Sect.~\ref{method}, 
we will present the results of the observations in
Sect.~\ref{results}, and discuss their implications in Sect.~\ref{discussion}.
In the following we use $H_0=65$~km~s$^{-1}$~Mpc$^{-1}$ and $q_0=0.5$; 
the cluster is thus a distance of 100~Mpc and 1~arcmin corresponds to 28.7~kpc.
\section{Data Set and Data Preparation}
\label{data}
\begin{figure}[t]
\begin{center}
\hspace*{-0.8cm}\resizebox{3.5in}{!}{\plotone{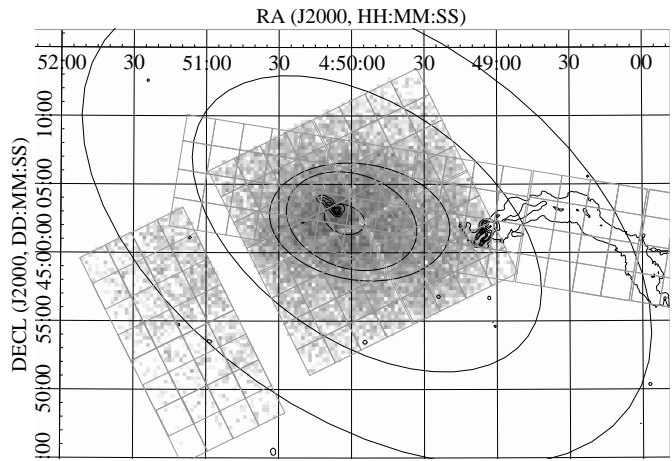}}
\end{center}
\caption{\label{f1} \small The ACIS~I image (0.5-7.5~keV)
with a 32$\times$32-pixel binning. 
The 4 ellipses are shown which have been used to characterize 
the symmetry of the ICM's surface brightness. The contour lines indicate the 
location of the radio galaxy 3C~129.1 at the cluster center and
3C~129 in the west of the cluster, and have been derived from 330~MHz VLA
observations (Lane et al. 2001). 
The small rectangles cover the ACIS~S and ACIS~I field of views
and show the sky-regions used for the deprojection analysis.
The deprojection analysis uses only the regions with their centers within 
the outer ellipse.
}
\end{figure}
We use two 3C~129 pointings, one 30~ksec pointing with the ACIS~S CCDs 
taken on December 9th, 2000, and one 10~ksec pointing with the 
ACIS~I CCDs taken on January 9th, 2001.
We detect X-ray emission near the radio cores of the radio galaxies 
3C~129 and 3C~129.1 and exclude the corresponding sky regions, 
as well as an additional point source in the south-east of the cluster center 
from the following analysis.
The radio galaxy 3C~129 has a 15~arcmin long tail which covers a substantial
fraction of the field of view of the first observation at radial distances 
from 10~arcmin to 25~arcmin from the cluster center. 
We exclude the corresponding sky region from the analysis.

The analysis is performed with the \verb+CIAO 2.1+ software. 
We divide the 8.4'$\times$8.4' solid angle regions of the sky covered by each 
ACIS chip in 4$\times$4 ``analysis regions'' of each 126''$\times$126'' 
(256$\times$256 pixels) solid angle coverage. 
This choice is a compromise between keeping the analysis feasible and using 
detector response functions which are as local as possible.
We construct background data sets with the tool \verb+make_acisbg+
using the blank field background data sets of M.~Markevitch.
Only events in the energy range from 1.2~keV to 7.5~keV are used.
We use a rather high lower threshold to minimize the influence
of uncertainties in the detector response and in the contribution of
the diffuse galactic X-ray emission. Above 7.5~keV the signal to noise 
ratio is poor. Application of the same requirements on the background 
rate as used 
for the production of the background data sets, reduces 
the exposure time of the first and second pointing to 19.5~ksec 
and 9.3~ksec, respectively.
We use the tools \verb+calcarf+ and \verb+calcrmf+ of A.~Vikhlinin to
compute for each analysis region an Auxiliary Response File (ARF) and a 
Response Matrix File (RMF) averaged over the corresponding detector area.
The analysis uses the ACIS~S chips S2-S4, and the ACIS~I chips I0-I4.
Figure \ref{f1} shows the binned ACIS~I picture together with the 
ACIS~S and ACIS~I fields of view, the analysis regions, 
and the location of the two radio sources.

\section{Spectroscopic Deprojection Analysis}
\label{method}
The iso-brightness contours of 3C~129 show a slightly elliptical 
shape with the major axis along the east-west direction.
Using the Chandra data together with the Einstein IPC contour map of 
\cite{Leah:00}, we define 4 ellipses with different centers and 
elasticities to parameterize the ICM surface brightness distribution 
at different distances from the cluster center (see Fig.~\ref{f1}).
Assuming rotational symmetry around the major axis of each ellipse
and using an interpolation scheme we define 11 approximately
ellipsoidal ICM shells. For each shell $s$ ($s=$1, 2, ..., 11)
we compute the volumes $V(s,r)$ along the line of sight of 
the 128 $\times$ 128 pixel analysis regions $r$
($r=$1, 2, ..., 95) by simple numerical integration.
For each shell we define one Raymond-Smith plasma model described 
by one set of parameters temperature, $T(s)$, particle density, $N(s)$,
and metal abundances, $A(s)$.
The algorithm works from shell to shell inwards, 
fitting the plasma parameters of the shell under consideration 
and using for the outer shells the best fit parameters determined before.
The expression minimized by the fit algorithm for shell $s$ (counting from
the cluster center outwards) reads:
\[
\chi^2(T(s),N(s),A(s);s)
)
\,\,=\,\,\rule{2.5cm}{0.0cm} 
\]
\[
\sum_{r' \in s}\,
\sum_{I'}\,
\left(\!C(r',I')-
\,
\frac{c(r')}{4\pi D_L\,^2}
\sum_{s'=s}^{11}\!
\int_0^\infty\!
V(s',r')\;
\times
\rule{0cm}{0.55cm} 
\right.
\]
\[
\kappa(N_{\rm H}(r');E')\,\,
j_{\rm RS}\left(T(s'),N(s'),A(s');E'\right)\,\times
\]
\begin{equation}
\left. 
D_{\rm r'}(I',E')\,dE'\rule{0cm}{0.55cm}\right)^2
\end{equation}
The first sum of the right term runs over the regions $r'$ which have their 
centers within the shell $s$ under consideration;
the second sum runs over the energy bins $I'$ of each energy spectrum; 
in the third sum $s'$ runs over the shell under consideration $s$ and all 
outer shells.
$C(r',I')$ is the number of counts found in energy bin $I'$ of 
region $r'$. The constants $c(r')\,\approx\,1$ allow for a slight deviation 
of the emissivity from the value expected from the assumed 
ellipsoidal geometry.
We adjust these constants after fitting the flux normalization of a shell.
The value $D_{\rm L}$ denotes the luminosity distance 
and $\kappa$ and $j_{\rm RM}$ describe 
the photo-electric absorption and the emissivity of the 
Raymond-Smith plasma as function of photon energy $E'$, respectively.
Finally, $D_{\rm r'}(I',E')$ is the detector response matrix
which gives the detection area times the probability with which an X-ray 
photon of energy $E'$ will be reconstructed within the energy bin $I'$
(averaged over the detector coordinates of region $r'$).
Due to the location of the cluster near the galactic plane we anticipated
substantial variation of the neutral hydrogen column density across the
field of view. We therefore performed radio observations with the Dominion
Astrophysical Observatory to obtain $N_{\rm H}$-values with 
arcminute resolution (principal investigator: T.\ Willis). 
Although the $N_{\rm H}$-values change by 25\% over the 
1.25$^\circ$$\times$2.5$^\circ$ field of view, 
the Chandra observations lie in a region of rather constant values 
of between 0.84$\times$$10^{22}$~cm$^{-2}$ and 0.9$\times$$10^{22}$~cm$^{-2}$. 
We use in the following the $N_{\rm H}$-values
from the radio observations. Independent determination of the mean 
$N_{\rm H}$-value from Chandra data gives consistent results.
Note that our inferred hydrogen column densities deviate from those 
inferred by Leahy et al. (2000) 
($N_{\rm H}$ = (0.57$\pm$0.03)$\times$$10^{22}$~cm$^{-2}$) 
from fitting the combined ROSAT and EXOSAT data with a one-temperature 
Raymond-Smith model.
For each shell $s$ we quote as radial distance $R_s$ the mean 
distances of all analysis regions contributing to the plasma fit of this shell.
We define the distance of a point from the cluster center
as the half-diameter of the major axis of the associated ellipse.

Technically the deprojection-analysis is performed with the 
\verb+CIAO+-tool \verb+sherpa+ running on a single 
script which contains all the information about the
geometry of the shells and analysis regions.
All errors are quoted on 90\% confidence level.
We estimate the increase of the statistical error on the parameters
of the inner shells due to the statistical uncertainty of the fit-parameters
of the outer shells with a simple Monte Carlo simulation and 
accordingly scale the errors on the fit-parameters.
The uncertainty of the background normalization is estimated to be 10\%
\cite{Vikh:01}. Determining the associated error in the parameters by 
repeating the full deprojection analysis with a background scaled 
up and down by 10\%, we add these errors in quadrature to the
statistical ones. 
The results of the outermost ICM shell depend on the assumed
shell thickness. Since this is an inherent limitation of the deprojection 
method we do not show the fit results of the 11th ICM shell in the 
following.
The fits give satisfactory $\chi^2$-values for 95 out of the 104 
analysis regions. For 9 regions large $\chi^2$-values indicate a deviation 
of the plasma properties from the ellipsoidal shell or contamination
by a field source, and we exclude these regions from the overall fit. 
\section{Results}
\label{results}
The Chandra image shows statistically significant evidence for X-ray emission 
out to 19 arcmin from the cluster core. We do not find evidence for
abrupt changes in the ICM surface brightness, so we have no direct evidence
for shocks or contact discontinuities.

\begin{figure}[th]
\begin{center}
\resizebox{2.9in}{!}{\plotone{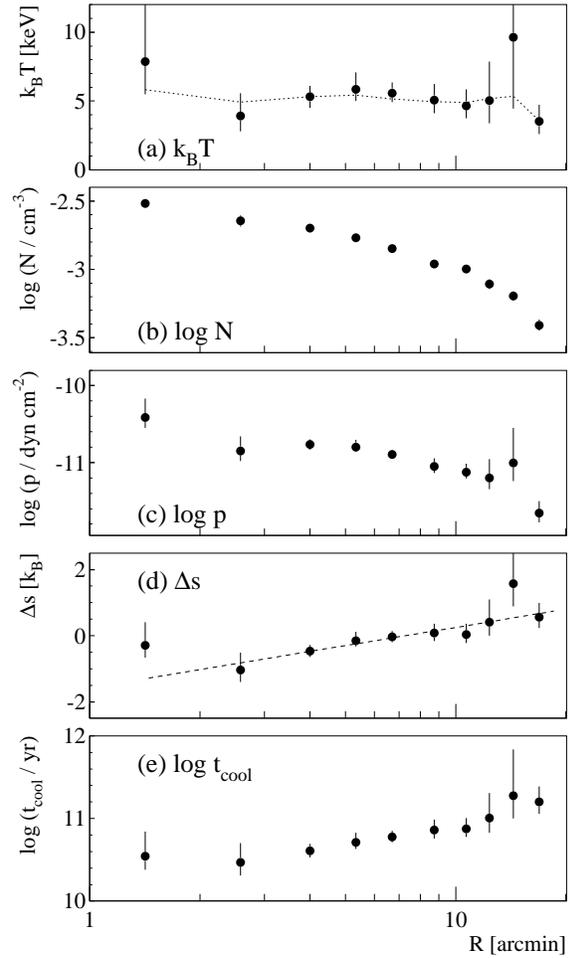}}
\vspace*{-0.5cm}
\end{center}
\caption{\label{f2} \small Results of the deprojection
analysis showing from top to bottom:
(a) temperature $T$, (b) particle density $N$, 
(c) gas pressure, (d) entropy, and (e) radiative cooling times.
The dotted line in panel (a) shows the projected temperature, 
and the dashed line in panel (d) is for illustrative purposes only
(see text).}
\vspace*{-0.5cm}
\end{figure}

Figure \ref{f2} shows the results of the deprojection analysis.
Averaged over all shells, the mean cluster temperature is 5.1$\pm$0.3~keV.
We have weak evidence for a hotter cluster core with 
$k_{\rm B}T\,=$ 7.9$^{+4.4}_{-2.4}$~keV at $R_1=$ 1.4~arcmin and 
a temperature decrease at the outermost 10th shell with
$k_{\rm B}T\,=$ 3.5$^{+1.2}_{-0.9}$~keV $R_{10}=$ 18.1~arcmin
($k_{\rm B}$ is the Boltzmann constant).
Various estimates show that the low temperature of the outermost shell
might be an artifact due to the soft diffuse galactic X-ray emission
which begins to substantially contaminate the cluster emission at these
distances from the cluster center.
The dotted line of Fig.~\ref{f2}(a) shows the projected temperatures 
determined by fits of one component Raymond-Smith models.
Naturally, the deprojected temperature varies more than the projected 
temperature which averages over several ICM shells.

The particle density (Fig.~\ref{f2}(b)) decreases monotonically
from 3.1$\times10^{-3}$ cm$^{-2}$ at the cluster center
to 3.9$\times10^{-4}$ cm$^{-3}$ at the cluster periphery.
The errors on the metal abundances of individual ICM shells are rather large
(typically about 0.4 solar abundances) and we do not show them here.
The mean value averaged over all shells is 0.2$\pm0.1$ solar abundance.
The thermal gas pressure, defined as $p\,$= $N\,k_{\rm B}\,T$ 
is shown in Fig.~\ref{f2}(c). 
The entropy per particle relative to the mean value of all shells 
$s_0$ is shown in Fig.~\ref{f2}(d) and has been computed 
from $s=$ $\frac{3}{2}$ $k_{\rm B}$ $\ln{(T\,\rho^{-2/3})}$ -
$s_0$ with $\rho\approx$ $0.6$ $m_{\rm P}$ $N$ and $m_{\rm P}$ 
the proton mass. The entropy profile will be discussed in more detail in
the next section.
Finally, the radiative cooling times are shown in
Fig.~\ref{f2}(e). 
The cooling times have been computed according to
$t_{\rm cool}\,=$  $\frac{3}{2}$ $N$ $k_{\rm B}$ $T$
$(\Lambda(N,T))^{-1}$ with the cooling function $\Lambda$ which denotes the
plasma emissivity including thermal bremsstrahlung and 
line emission processes \cite{Suth:93}.
All radiative cooling times exceed the Hubble time, the 
shortest cooling time being 30$^{+20}_{-9}$~Gyr.
\section{Discussion}
\label{discussion}
Based on a deprojection analysis of ROSAT surface brightness data 
Leahy \& Yin (2000) inferred a cluster temperature decreasing from
$\simeq$6~keV at the cluster periphery (at $\simeq$20 arcmin from the 
cluster core) to about $\simeq$3~keV at the cluster center.
The Chandra imaging spectroscopy data however, does not confirm this
temperature decrease. On the contrary, both, the projected and deprojected 
temperature profiles indicate a hot cluster core.
The discrepancy might be explained by one or more of the numerous assumptions 
entering the deprojection of surface brightness images mentioned in the 
introduction, as e.g.\ a true dark matter core radius substantially deviating
from the assumed one.

A hot cluster core might be the result of shock heating. 
This possibility can be studied with the entropy per particle which
is able to distinguish between ICM heating by adiabatic compression or 
by shocks. By definition, adiabatic compression does not increase the 
ICM entropy, while shock heating, converting the energy of bulk plasma 
motion into heat, does increase it.
A possible interpretation of the entropy profile is based on the dashed line
in Fig.~\ref{f2}(d) which shows a fit to the entropy of the second to
tenth shell and suggests that the entropy increases with distance 
from the cluster center. Indeed, in the framework of hierarchical clustering 
scenarios one expects that cluster formation through mass accretion 
produces a radially increasing ICM entropy as consequence of the 
growing strength of the cluster accretion shock as the total virialized 
cluster mass accumulates (see e.g.\ Tozzi et al.\ 2001, and references 
therein).
The large entropy of the cluster core might be the result of shock heating
by a jet or wind from the central radio galaxy 3C~129.1.
The required power is modest: $\simeq 10^{43}$ 
$\frac{\Delta (k_{\rm B}\,T)}{\rm 5\,keV}$
$\frac{t_{\rm 3C129.1}}{\rm 100~Myr}$ ergs s$^{-1}$
where $\Delta (k_{\rm B}\,T)$ is the temperature increase at the 
cluster center and $t_{\rm 3C129.1}$ denotes the lifetime of the source.
However, at the cluster center we do not find structure in the ICM surface 
brightness in direct support of this hypothesis.

The analysis presented here is plagued by large statistical
and systematic errors. The first stem from the modest integration time
and the low surface brightness of 3C~129. 
The latter mainly results from the asymmetric shape of the ICM and possible
contribution of the diffuse galactic X-ray emission at larger cluster core 
distances.
For brighter, more symmetric clusters the spectroscopic deprojection technique 
opens the possibility to determine the shape of the dark matter potential.
The only uncertainties of such an analysis derive from the uncertain 
ICM geometry and from undetected ICM pressure components as kinetic pressure,
magnetic field pressure and Cosmic Ray 
pressure. 
\\[2ex]
{\it Acknowledgments:} The author would like to thank H.\ V\"olk and 
D.\ Harris for fruitful discussions on X-ray emission from galaxy cluster 
and AGN jets. W.M.~Lane kindly provided the 330~MHz VLA map shown in 
Fig.~\ref{f1}, and T.~Willis the $N_{\rm H}$-data used for the deprojection 
analysis. Very helpful comments by an anonymous referee are gratefully
acknowledged. 
This research has been supported by the NASA (NAS8-39073 and GO 0-1169X). 


\begin{thebibliography}{}
{\small
\bibitem[Allen et al.\ 2001]{Alle:01}{Allen, S.W., Ettori S., Fabian A.C.\
2001, MNRAS , 324, 877}
\bibitem[Arnaud et al.\ 2000]{Arna:00}{Arnaud, M., Aghanim, N., Gastaud, R., 
et al.\ 2000, A\&A, 365, L67}
\bibitem[Edge \& Stewart 1991]{Edge:91}{Edge, A.C., Stewart, G.C.\ 1991,
MNRAS, 252, 414}
\bibitem[Evrard 1990]{Evra:90}{Evrard, A.E.\ 1990, ApJ, 363, 349}
\bibitem[Eilek 1999]{Eile:99}{Eilek, J.\ 1999, MPE Report 271, 71}
\bibitem[Fabian 1994]{Fabi:94}
{Fabian, A.\ C.\ ARAA, 1994, 32, 277}
\bibitem[Honda et al.\ 1996]{Hond:96}{Honda, H., et al. 1996, ApJ, 473, L71}
\bibitem[Kaastra et al.\ 2001]{Kaas:01}{Kaastra, J.S., Ferrigno, C.,
Tamura, T., et al.\ 2001, A\&A, 365, L99}
\bibitem[Kikuchi et al. 1999]{Kiku:99}{Kikuchi, K., Furusho, T., Ezawa, H., 
Yamasaki, N., Ohashi, T., Fukazawa, Y., Ikebe, Y.\ 1999, PASJ, 51, 301}
\bibitem[Lane et al.\ 2001]{Lane:01}{
Lane, W.M., Harris, D.E., Ensslin, T.A., Kassim, N.E., Perley, R.A.\
2001, AAS, 199, 9814}
\bibitem[Leahy \& Yin 2000]{Leah:00}{Leahy, D.A., Yin, D.\ 2000,
MNRAS, 313, 617}
%
\bibitem[Markevitch et al.\ 1998]{Mark:98} {Markevitch, M., 
Forman, W.R., Sarazin, C.L., Vikhlinin, A.\ 1998, ApJ, 503, 77}
\bibitem[Markevitch et al.\ 2000]{Mark:00a} {Markevitch, M., Ponman,
T.J., Nulsen, P.E., et al.\ 2000, ApJ, 541, 542}
\bibitem[Metzler \& Evrard 1994]{Metz:94}
{Metzler, C.A., Evrard, A.E.\ 1994, ApJ, 437, 564}
\bibitem[McNamara et al.\ 2000]{McNa:00} 
{McNamara, B.R., Wise, M., Nulsen, P.E.J., et al.\ 2000, ApJ, 534, L135}
\bibitem[Narayan \& Medvedev 2001]{Nara:01}{
Narayan, R., Medvedev, M.V., ApJ, 562, L129}
\bibitem[Roettiger et al.\ 1996]{Roet:96} 
{Roettiger, K., Burns, J.O., Loken, C.\ 1996, ApJ, 473, 651}
\bibitem[Schmidt et al.\ 2001]{Schm:01}{Schmidt, R.W., Allen S.W., Fabian A.C.\
 2001, MNRAS, 327, 1057}
\bibitem[Sutherland \& Dopita 1993]{Suth:93} {Sutherland, R.S., Dopita, M.A.\
1993, ApJS, 88, 235}
\bibitem[Tamura et al.\ 2001]{Tamu:01}
{Tamura, T., Kaastra, J.S., Peterson, J.R., et al.\ 2001, A\&A, 365, L87}
\bibitem[Tozzi \& Norman 2001]{Tozz:01}
{Tozzi, P., Norman, C.\ 2001, ApJ, 546, 63}}
\bibitem[Vikhlinin et al.\ 2001]{Vikh:01}{Vikhlinin, A., Markevitch, M., Murray, S.S.\ 2001, ApJ, 551, 160}
\bibitem[White 2000]{Whit:2000}{White, D. A. 2000, MNRAS, 312, 663}
\end{thebibliography}
\end{document}